\documentclass[aps,pre,twocolumn,groupedaddress]{revtex4}

\usepackage{graphicx}
\usepackage{dcolumn}
\usepackage{bm}
\usepackage{subfigure}

\begin{document}


\title{Behavior of the Random Field $XY$ Model on Simple Cubic Lattices at $h_r =~1.5$}


\author{Ronald Fisch}
\email[]{ronf124@gmail.com}
\affiliation{382 Willowbrook Dr.\\
North Brunswick, NJ 08902}


\date{\today}

\begin{abstract}
We have performed studies of the 3D random field $XY$ model on 32 samples of
$L \times L \times L$ simple cubic lattices with periodic boundary conditions,
with a random field strength of $h_r$ = 1.5, for $L =$ 128, using a parallelized
Monte Carlo algorithm.  We present results for the sample-averaged magnetic
structure factor, $S (\vec{\bf k})$ over a range of temperature, using both
random hot start and ferromagnetic cold start initial states, and $\vec{\bf k}$
along the [1,0,0] and [1,1,1] directions.  At $T =$ 1.875, $S (\vec{\bf k})$
shows a broad peak near $|\vec{\bf k}| = 0$, with a correlation length which is
limited by thermal fluctuations, rather than the lattice size.  As $T$ is
lowered, this peak grows and sharpens.  By $T =$ 1.5, it is clear that the
correlation length is larger than $L =$ 128.  The lowest temperature for which
$S (\vec{\bf k})$ was calculated is $T =$ 1.421875, where the hot start and cold
start initial conditions are usually not finding the same local minimum in the phase
space.  Our results are consistent with the idea that there is a finite value of $T$
below which $S (\vec{\bf k})$ diverges slowly as $|\vec{\bf k}|$ goes to zero.  This
divergence would imply that the relaxation time of the spins is also diverging.  That
is the signature of an ergodicity-breaking phase transition.

\end{abstract}


\maketitle

\section{Introduction}

The behavior of the three-dimensional (3D) random-field $XY$ model
(RFXYM) at low temperatures and weak to moderate random field
strengths continues to be controversial.  A detailed calculation by
Larkin\cite{Lar70} showed that, in the limit that the number of
spin components, $n$, becomes infinite, the ferromagnetic phase
becomes unstable when the spatial dimension of the lattice is less
than or equal to four, $d \le 4$.  Dimensional reduction
arguments\cite{IM75,AIM76} appeared to show that the long-range order
is unstable for $d \le 4$ for any finite $n \ge 2$.  However, there
are several reasons for questioning whether dimensional reduction
can be trusted for $XY$, {\it i.e.} $n = 2$, spins.

The existence of replica-symmetry breaking (RSB) in random field models
was first shown by Mezard and Young\cite{MY92} in 1992.  Mezard and
Young emphasized the Ising case, and the fact that this applies for all
finite $n$ seems to have been overlooked by most people for a number
of years.  The result was confirmed by Brezin and De Dominicis,\cite{BDD01}
who also emphasized the Ising case. A detailed analysis of
perturbation theory finds that dimensional reduction is not correct.
The renormalization group critical point describing the paramagnet to
ferromagnet phase transition becomes unstable in six dimensions.
They argue that below six dimensions there is a phase transition from
the paramagnetic phase into a RSB glassy phase which
has no magnetization.  It is expected that there is still a ferromagnetic
phase below the glassy phase for some range of dimensions below six, but
this point is not discussed in detail.

Some time ago, Monte Carlo calculations\cite{GH96,Fis97} showed that
there was a line in the temperature vs. random-field plane of the phase
diagram of the three-dimensional (3D) random-field $XY$ model (RFXYM),
at which the magnetic structure factor becomes large as the wave-number
$k$ becomes small.  Gingras and Huse\cite{GH96} claim that the phase
transition occurs at the temperature where vortex lines undergo a
percolation transition, as is true for the pure 3D $XY$ model.
The current author does not understand why this should be an exact
result when there is a random field, but it seems to be a good
approximation.  Additional calculations\cite{Fis07} indicated that
there appeared to be small jumps in the magnetization and the energy of
$L$ = 64 lattices at a random field strength of $h_r = 2.0$, at a
temperature somewhat below $T = 1.0$.  Further calculations\cite{Fis10}
showing similar behavior for other values of the random field strength
were also performed. If such behavior persisted for larger values of $L$,
with the sizes of these jumps being independent of $L$ for large $L$,
this would demonstrate that there is a ferromagnetic phase at weak to
moderate random fields and low temperatures for this model.  However,
Aizenman and Wehr\cite{AW89,AW90} have proven under certain conditions
that this should not happen in 3D.  The sizes of these jumps should
scale to zero as $L$ goes to infinity.  The rates of the scaling
characterizes the phase transition, analogous to the critical exponents
which describe critical behavior in second order phase transitions.
Behavior of this type would appear to be a reasonable description of
the phase transition from the paramagnet to the RSB phase predicted by
Brezin and De Dominicis\cite{BDD01}  This type of behavior was recently
seen in Monte Carlo calculations by the author\cite{Fis19} at $h_r = 1.875$.

The work reported here describes Monte Carlo calculation conducted at a
random field strength of $h_r = 1.5$.  The results for $L \times L \times L$
simple cubic lattices with $L = 128$ will be presented.  One significance of
$h_r = 1.5$ is that Garanin, Chudnovsky and Procter\cite{GCP13} have
claimed that in the 3D RFXYM there is a large magnetization at $T = 0$ for
this value of $h_r$.  The region of the phase diagram which is studied here
also overlaps the region studied by Gingras and Huse.\cite{GH96}

\section{The Model}

For fixed-length classical spins the Hamiltonian of the RFXYM is
\begin{equation}
  H ~=~ - J \sum_{\langle ij \rangle} \cos ( \phi_{i} - \phi_{j} )
  ~-~ h_r \sum_{i} \cos ( \phi_{i} - \theta_{i} )  \, .
\end{equation}
Each $\phi_{i}$ is a dynamical variable which takes on values
between 0 and $2 \pi$. The $\langle ij \rangle$ indicates here a sum
over nearest neighbors on a simple cubic lattice of size $L \times L
\times L$. We choose each $\theta_{i}$ to be an independent
identically distributed quenched random variable, with the
probability distribution
\begin{equation}
  P ( \theta_i ) ~=~ 1 / 2 \pi   \,
\end{equation}
for $\theta_i$ between 0 and $2 \pi$.  We set the exchange constant to
$J = 1$.  This gives no loss of generality, since it merely defines the
temperature scale.  This Hamiltonian is closely related to models of
vortex lattices and charge density waves.\cite{GH96,Fis97}

Larkin\cite{Lar70} studied a model for a vortex lattice in a
superconductor.  His model replaces the spin-exchange term of the
Hamiltonian with a harmonic potential, so that each $\phi_{i}$ is no
longer restricted to lie in a compact interval.  He argued that for
any non-zero value of $h_r$ this model has no ferromagnetic phase on
a lattice whose dimension $d$ is less than or equal to four.  The Larkin
approximation is equivalent to a model for which the number of spin
components, $n$, is sent to infinity.  A more intuitive derivation
of this result was given by Imry and Ma,\cite{IM75} who assumed that
the increase in the energy of an $L^d$ lattice when the order parameter
is twisted at a boundary scales as $L^{d - 2}$ for all $n > 1$, just as
it would for $h_r = 0$.  Using this assumption, they argued that when
$d \le 4$ there is a length $\lambda$, now called the Imry-Ma length, at
which the energy which can be gained by aligning a local spin domain with
its local random field exceeds the energy cost of forming a domain wall.
They claimed that this implies the magnetization would decay to zero when
the system size, $L$, exceeds $\lambda$.

Within a perturbative $\epsilon$-expansion one finds the phenomenon
of ``dimensional reduction"\cite{AIM76} for the properties of the
paramagnetic-to-ferromagnetic critical point.  The critical exponents of
any $d$-dimensional $O(n)$ random-field model appear to be identical to
those of an ordinary $O(n)$ model of dimension $d - 2$. For the $n = 1$
(RFIM) case, this was soon shown rigorously to be incorrect for
$d < 4$.\cite{Imb84,BK87}  However, Brezin and De Dominicis\cite{BDD01}
later showed that the existence of RSB in this model\cite{MY92} means that
the paramagnetic-to-ferromagnetic critical point is unstable in less than
six dimensions.  More recently, extensive numerical results for the Ising
case at $T = 0$ have been obtained for $d = 4$ and $d=5$.\cite{FMPS17a,FMPS17b}
They determined that dimensional reduction is ruled out numerically in the
Ising case for $d = 4$, but not for $d = 5$.\cite{FMPS18}  The algorithm
used to obtain these numerical results for the RFIM does not work for $T > 0$,
and it is not clear what the finite $T$ behavior should be.  According to
Brezin and De Dominicis,\cite{BDD01} there should be a glassy RSB phase
sandwiched between the paramagnet and the ferromagnet when $d < 6$.  This
behavior is likely to occur in the RFXYM also, as long as $d$ is high enough
for a ferromagnetic phase to exist.  Further, there does not seem to be any
reason why a glassy phase should not continue to exist for the RFXYM in
$d = 3$, even if there is no ferromagnetic phase.

The scaling behavior at low $T$ is somewhat different for $n \ge 2$.  Because
translation invariance is broken for any non-zero $h_r$, it seems quite
implausible to the current author that the twist energy for Eqn.~(1)
scales as $L^{d - 2}$ for large $L$ when $d \le 4$, even though this is
correct to all orders in perturbation theory.  The problem with assuming
this scaling is that the Irmy-Ma length provides a natural length scale
to the problem. We need to scale out to the Imry-Ma length before we can
learn the true long-distance behavior of the model.  This means that the
effective strength of the randomness cannot be assumed to grow without
bound when $d \le 4$, just because it grows for weak non-zero $h_r$.
We must do an detailed calculation to find out what actually happens.

This point needs to be emphasized.  When the random field is weak, the
Imry-Ma length, $\lambda$, becomes long.  No matter how weak the random
field is, we must always go to lengths larger than $\lambda$ to see the
crossover to the true thermodynamic limit.  In this work we will
demonstrate numerically that the calculations of Gingras and Huse\cite{GH96}
were done on lattices which were too small to reveal this true
thermodynamic limit.  This is also true of the current author's work
done on the model at that time.\cite{Fis97}

An alternative derivation of the Imry-Ma result by Aizenman and
Wehr,\cite{AW90} which claims to be mathematically rigorous,
also makes an assumption that the model is defined on a lattice
which has a probability distribution which is invariant under
rotation and translation.  Thus, their argument is only rigorous
for a model which is defined on some lattice which is locally
disordered, but has rotational invariance on the average.

It may be that there exists a better argument, which can show that
this technical issue is not essential.  It is not clear, however,
such an argument ought to exist.  It could be true that, in the 3D
$n = 2$ case, the Imry-Ma argument fails when the random fields
are weak enough, as a consequence of the existence of vortex lines
on the dual lattice.  This possibility has been suggested by a number
of authors, e.g. Chudnovsky and coworkers.\cite{GCP13,PGC14}
However, the current author does not find the existing numerical
work by the Chudnovsky group to be convincing, because they are not
using weak random fields.

The model we study here is defined on a finite simple cubic lattice,
which does not have the property of average rotational invariance.
Although the average over the probability distribution of random
fields restores translation invariance, one must take the infinite
volume limit first.  It is not correct to interchange the infinite
volume limit with the average over random fields.  Taking an average
over random field configurations does not remove the necessity of
going beyond the Imry-Ma length to reach the large system behavior.

This problem of the interchange of limits is equivalent to the
existence of RSB.  A functional renormalization group calculation
going to two-loop order was performed by Tissier and Tarjus,\cite{TT06}
and independently by Le Doussal and Wiese.\cite{LW06}  They found
that there was a stable critical fixed point of the renormalization
group for some range of $d$ below four dimensions in the $n = 2$
random field case.  However, it is not clear from their calculation
what the nature of the low-temperature phase is, or whether this fixed
point is stable down to $d = 3$.  Tarjus and Tissier\cite{TT08}
later presented an improved version of this calculation, which explains
more explicitly why dimensional reduction fails for the $n = 2$ case
when $d \leq 4$.  The difference between these calculations and the RSB
calculations is that they are looking at the stability of the
ferromagnetic phase near $T = 0$, and not the stability of the
paramagnet-ferromagnet transition.

\section{Structure factor and magnetic susceptibility}

The magnetic structure factor, $S (\vec{\bf k}) = \langle
| \vec{\bf M}(\vec{\bf k}) |^2 \rangle $, for $XY$ spins is
\begin{equation}
  S (\vec{\bf k}) ~=~  L^{-3} \sum_{ i,j } \cos ( \vec{\bf k} \cdot
  \vec{\bf r}_{ij}) \langle \cos ( \phi_{i} - \phi_{j}) \rangle  \,   ,
\end{equation}
where $\vec{\bf r}_{ij}$ is the vector on the lattice which starts
at site $i$ and ends at site $j$, and here the angle brackets denote
a thermal average.  For a random field model, unlike a random bond
model, the longitudinal part of the magnetic susceptibility, $\chi_{||}$,
which is given by
\begin{equation}
  T \chi_{||} (\vec{\bf k}) ~=~ 1 - M^2 ~+~ L^{-3} \sum_{ i \ne j } \cos (
  \vec{\bf k}  \cdot \vec{\bf r}_{ij}) (\langle \cos ( \phi_{i} - \phi_{j}
  ) \rangle ~-~ Q_{ij} )  \,   ,
\end{equation}
is not the same as $S (\vec{\bf k})$ even above $T_c$.  For $XY$ spins,
\begin{equation}
  Q_{ij} ~=~ \langle \cos ( \phi_{i} ) \rangle \langle \cos (
  \phi_{j} ) \rangle ~+~ \langle \sin ( \phi_{i} ) \rangle \langle
  \sin ( \phi_{j} ) \rangle  \,  ,
\end{equation}
and
\begin{equation}
  M^2 ~=~ L^{-3} \sum_{i} Q_{ii}
      ~=~ L^{-3} \sum_{i} [ \langle \cos ( \phi_{i} ) \rangle^2 ~+~
       \langle \sin ( \phi_{i} ) \rangle^2 ] \,  .
\end{equation}
When there is a ferromagnetic phase transition, $S ( \vec{\bf k} = 0 )$
has a stronger divergence than $\chi ( \vec{\bf k} = 0 )$.

The scalar quantity $\langle M^2 \rangle$, when averaged over a set
of random samples of the random fields, is a well-defined function
of the lattice size $L$ for finite lattices. With high probability,
it will approach its large $L$ limit smoothly as $L$ increases.
The vector $\vec{\bf M}$, on the other hand, is not really a
well-behaved function of $L$ for an $XY$ model in a random field.
Knowing the local direction in which $\vec{\bf M}$ is pointing,
averaged over some small part of the lattice, may not give us a
strong constraint on what $\langle \vec{\bf M} \rangle$ for the
entire lattice will be.  When we look at the behavior for
all $\vec{\bf k}$, instead of merely looking at $|\vec{\bf k}| = 0$,
we get a much better idea of what is really happening.

\section{Numerical results for $S ( \vec{\bf k} )$ and $\chi (|\vec{\bf k}| = 0)$}

In this work, we will present results for $S (\vec{\bf k})$.  The data
were obtained from $L \times L \times L$ simple cubic lattices with
$L = 128$ using periodic boundary conditions.  The calculations
were done using a clock model which has 12 equally spaced dynamical
states at each site.  In addition, there is a static random phase at
each site which was chosen to be $0, \pi/24, \pi/12$ or $3\pi/24$ with
equal probability.  This random phase does not play an essential role,
but it is convenient.  It reduces the effective strength of the 12-fold
anisotropy without a significant slowing down of the computer algorithm.
It is expected to reduce the chance of any issue with the quality of the
pseudorandom-number generators.  It also provides an increased number of
ordered initial states for the calculations which start in such ordered
initial states.  The algorithm used in this work is a version of the
algorithm which was used in our earlier calculations.\cite{Fis19}

The idea of adding $p$-fold symmetry-breaking terms to an $XY$ model
goes back to Jose, Kadanoff, Kirkpatrick and Nelson,\cite{JKKN77} who
studied the effects of nonrandom fields of this type on the
Kosterlitz-Thouless (KT) transition in 2D.  The result they found was
that the KT transition survives the addition of terms of this type near
$T_c$ if $p > 4$, but that the system becomes ferromagnetic at some
lower value of $T$.  This work was extended to $p$-fold fields which
varied randomly in space by Houghton, Kenway and Ying\cite{HKY81} and
Cardy and Ostland.\cite{CO82}  It was found that the KT transition
survives in the random $p$-fold field case for $p \ge 3$.

Generalizing this idea to $d > 2$ is straightforward.  It has been
known for some time that a nonrandom $Z_p$ model of this type is in the
universality class of the ferromagnetic $XY$ model whenever $p > 4$.\cite{WK74}
For random phase $Z_p$ models without a random-field term, there are no
analytical results.  However, it has been found numerically that in 3D the
model is in the universality class of the pure $XY$ model under most
conditions, even if the number of dynamical states of each spin is only
3.\cite{Fis92}  Under conditions of very low temperature, this model may
undergo an incommensurate-to-commensurate type of charge-density wave phase
transition.  Thus it is expected that, when we include the random-field term,
the model will behave essentially as a random-field $XY$ model, as long as we
do not attempt to work at very low temperatures and random field strengths
much weaker than the ones used here.\cite{Fis97}  However, we want to have
more than merely being in the same universality class, which only requires 3
dynamical states at each site.  We have found that if we use at least 8
dynamical states at each site, then the results we find numerically do not
depend on the number of dynamical states, at least for $T \ge 1.00$.

Based on earlier Monte Carlo calculations,\cite{GH96,Fis07} we know the
approximate location of the phase boundary in the ($h_r , T$) plane.
This is true despite the fact that we are not certain what the nature of
the low temperature phase is.  The reason why this is possible is that
we are able to locate the phase boundary by finding where the static
ferromagnetic correlation length first diverges as we lower $T$ or $h_r$.
It was not known {\it a priori} if it would be possible to do calculations
under conditions where we could get past the crossover region and see the
large lattice behavior on the phase boundary.

The direction of the random field at site $i$, $\theta_i$, was chosen
randomly from the set of the 48th roots of unity, independently at each
site.  Since $\theta_i$ has 48 possible values, our past experience with
models of this type\cite{Fis19} indicates that there is no reason to expect
that the discretization will affect the behavior in an observable way.

The computer program uses three independent pseudorandom number generators:
one for choosing initial values of the dynamical variables, $\phi_i$, in
the hot start initial condition, one for setting the static random phases,
$\theta_i$, and a third one for the Monte Carlo spin flips, which are
performed by a single-spin-flip heat-bath algorithm.

The pseudorandom-number generators for the $\phi_i$ and the $\theta_i$ are
standard linear congruential generators which have been used for many years.
Given the same initial seeds, they will always produce the same string of
numbers, which is a property needed by the program.  They have excellent
statistical properties for strings of numbers up to length $10^8$ or so,
which is adequate for our purpose here.  Using separate generators for
choosing the initial values of the dynamical $\phi_i$ and the static random
$\theta_i$ was not really necessary, since the hot starts were always done
at a high value of $T$.  However, the cost of doing this is negligible, and
it would have allowed the use of random initial start conditions at any
value of $T$, although that was not done in the work reported here.

The pseudorandom-number generator used for the Monte Carlo spin flips
was the library function $random\_number$ supplied by the Intel
Fortran compiler, which is suitable for parallel computation.  It is
believed that this generator has good statistical properties for
strings of length $10^{14}$, which is what we need here.  However,
the author has no ability to check this for himself.  The spin-flip
subroutine was parallelized using OpenMP, by taking advantage of the
fact that the simple cubic lattice is two-colorable.  It was run on
Intel multicore processors of the Bridges Regular Memory machine at
the Pittsburgh Supercomputer Center.  The code was checked by setting
$h_r = 0$, and seeing that the known behavior of the pure ferromagnetic
3D XY model was reproduced correctly.  It was found, however, that
using more than two cores in parallel did not result in any additional
speedup of the calculation.  This made it impractical to study 3D
lattices larger than $L = 128$.

32 different realizations of the random fields $\theta_i$ were studied.
Each lattice was started off in a random spin state at $T = 2.375$, above
the $T_c$ for the pure $O(2)$ model, which is approximately 2.202.\cite{Jan90}
The $T_c$ for a pure $Z_4$ model is 2.2557, half that of the pure Ising model.
As far as the author knows, there are no highly accurate calculations of $T_c$
for pure $Z_p$ models with $p > 4$ on a simple cubic lattice.  It is expected,
however, that these will converge to the $T_c$ for the $O(2)$ model
exponentially fast in $n$.  The reason for this is that
$\cos (\theta_j - \theta_i)$ for nearest neighbor $i$ and $j$ at $T_c$, which
is the energy per bond at $T_c$, is 0.33 on this lattice.  This means that
the typical angle between nearest neighbor spins at $T_c$ is slightly less
than $2 \pi / 5$.  Once the mesh size for $\theta_i$ becomes less than the
typical value of $\theta_j - \theta_i$, the effect of the discretization
disappears rapidly.

Each lattice was then cooled slowly to $T = 1.421875$, using a cooling
schedule which depended on $T$.  Although the relaxation of the spins is
not a simple exponential function, it is quite apparent that the relaxation
is becoming very slow as $T = 1.421875$ is approached.  At $T = 1.421875$,
the sample was relaxed until an apparent equilibrium was reached over an
appropriate time scale.  This time scale was at least 737.280 Monte Carlo
steps per spin (MCS).  Some samples required relaxation for up to three
times longer than these minimum times.

\begin{figure}
\includegraphics[width=3.4in]{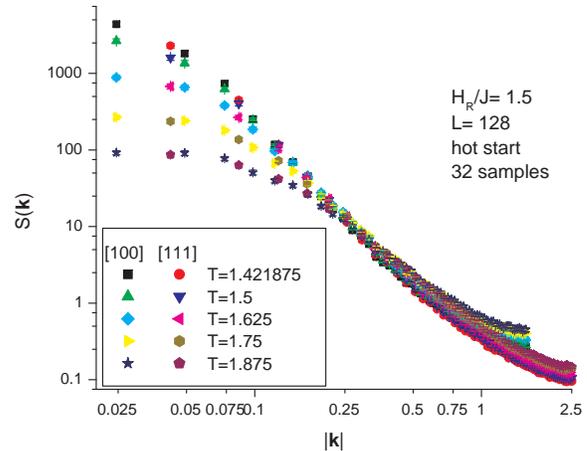}
\caption{\label{Fig.1}(color online) Structure factor vs. $|\vec{\bf k}|$ for
$128 \times 128 \times 128$ lattices with $h_r$ = 1.5 at various
temperatures, using slowly cooled spin states. Both the $x$-axis and the
$y$-axis are scaled logarithmically.  The points shown are averages of data
along [1,0,0] or [1,1,1] directions.  One $\sigma$ statistical errors are
approximately the size of the plotting symbols.}
\end{figure}

\begin{figure}
\includegraphics[width=3.4in]{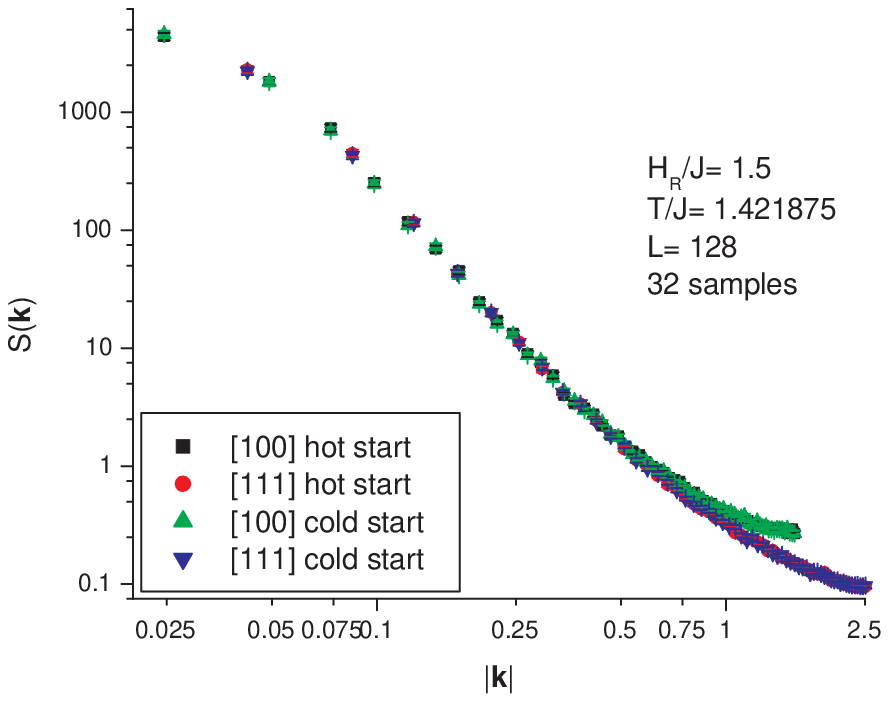}
\caption{\label{Fig.2}(color online) Structure factor vs. $|\vec{\bf k}|$ for
$128 \times 128 \times 128$ lattices with $h_r$ = 1.5 at $T =$ 1.421875,
comparing the slowly warmed states with the slowly cooled states.
The points shown are averages of data along [1,0,0] or [1,1,1] directions.
Both the $x$-axis and the $y$-axis are scaled logarithmically.  One $\sigma$
statistical errors are approximately the size of the plotting symbols.}
\end{figure}

\begin{figure}
\includegraphics[width=3.4in]{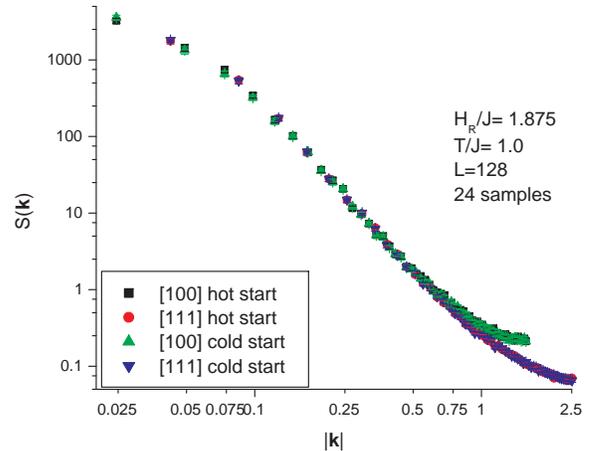}
\caption{\label{Fig.3}(color online) Structure factor vs. $|\vec{\bf k}|$ for
$128 \times 128 \times 128$ lattices with $h_r$ = 1.875 at $T =$ 1.0,
comparing the slowly warmed states with the slowly cooled states.
The points shown are averages of data along [1,0,0] or [1,1,1] directions.
Both the $x$-axis and the $y$-axis are scaled logarithmically.  One $\sigma$
statistical errors are approximately the size of the plotting symbols.  This
figure shows a reanalysis of data from ref.\cite{Fis19}.}
\end{figure}

After each sample was relaxed at $T$ = 1.421875, a sequence of 6
equilibrated spin states obtained at intervals of 40,960 MCS was Fourier
transformed and averaged to calculate $S (\vec{\bf k})$. Finally, an
average over the 32 samples was performed.  Similar procedures were
followed at higher values of $T$, where the equilibration times were
shorter.  The results for $S (\vec{\bf k})$ along the [1,0,0] and [1,1,1]
directions at a sequence of temperatures from $T = 1.875$ down to $T =
1.421875$ is shown in Fig.~1.  In this range of $T$, for small values of
$|\vec{\bf k}|$, $S (|\vec{\bf k}|)$ is increasing as $T$ is lowered.  At
$T = 1.875$, $S (|\vec{\bf k}|)$ is virtually independent of $|\vec{\bf k}|$
for small $|\vec{\bf k}|$, indicating that the spin correlations are
limited by thermal fluctuations.  At $T = 1.421875$, the spin correlations
continue to increase as $|\vec{\bf k}|$ gets smaller, indicating that the spin
correlation length is greater than the lattice size.  However, the small
$|\vec{\bf k}|$ data for $T = 1.421875$ do not fall on a straight line on
this log-log plot.  We do not know what would happen for larger lattices,
but we have no evidence that the data can be explained by a critical point
with a correlation length that diverges like some power of temperature.

Data were also obtained for the same sets of samples using ordered
initial states and warming from $T$ = 1.375.  At least two, and sometimes
more initial ordered states were used for each sample.  The initial
magnetization directions used were chosen to be close to the direction
of the magnetization of the slowly cooled sample with the same set of
random fields.  This type of initial state was chosen because it was
found in the earlier work\cite{Fis07} that this is the way to find the
lowest energy minima in the phase space.  The data from the initial
condition which gave the lowest average energy for a given sample was
then selected for further analysis and comparison with the slowly cooled
state data for that sample.  The relaxation procedure at $T = 1.421875$
for the warmed states was the same one used for the cooled states, and
the calculation of $S( |\vec{\bf k}| )$ proceeded in the same way.  In Fig.~2
we compare the $S( |\vec{\bf k}| )$ for the slowly warmed initial states
with the data for the slowly cooled initial states at $T = 1.421875$.

For purposes of comparison with Fig.~2, in Fig.~3 we show data from ref.\cite{Fis19},
analyzed in the same way.  These data for lattices with $h_r = 1.875$ at $T = 1.0$
are qualitatively similar to the data in Fig.~2.  Although $h_r$ is now larger, $T$
is smaller.  Thermal disorder in Fig.~2 is being replaced by random-field induced
disorder in Fig.~3.  The resulting change in $S( |\vec{\bf k}| )$ is not zero, but
it is small.  The crossover from the large $|k|$ behavior to the small $|k|$ behavior,
which happens at $1 / \lambda$, is at a somewhat larger value of $|k|$ in Fig.~3, as
predicted by the Imry-Ma argument.  Note that the original Imry-Ma argument\cite{IM75}
is a zero-temperature argument.  One should not assume this idea of thermal disorder
replacing random-field induced disorder will work for larger values of $n$, unless
and until some evidence of that is found.  The small $|k|$ region of Fig.~3 does not
appear to be approaching a finite value for $S$ in the limit $|k| \to 0$, as discussed
in more detail in ref.\cite{Fis19}.

The reader should note that the estimates of $\lambda$ from Fig.~2 and Fig.~3 imply
that the $L = 96$ lattices studied by Gingras and Huse\cite{GH96} are passing through
$\lambda$ close to $h_r = 1.3$, which is the point where Gingras and Huse claim a
phase transition occurs at $T = 1.5$.  This coincidence means that their ideas about
the nature of the phase transition are not reliable, because their lattices are not
large enough to have reached the true small $|k|$ region at their estimated value of
the phase transition point.

\begin{figure}
\includegraphics[width=3.4in]{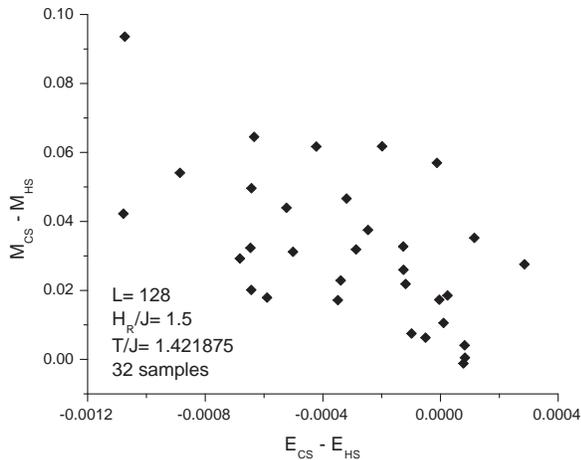}
\caption{\label{Fig.4} Jump in the magnetization vs. jump
in the energy for $128 \times 128 \times 128$ lattices with $h_r$ = 1.5
at $T$ = 1.421875.  States with hot start and ordered cold start initial
conditions are compared for each sample.}
\end{figure}

\begin{figure}
\includegraphics[width=3.4in]{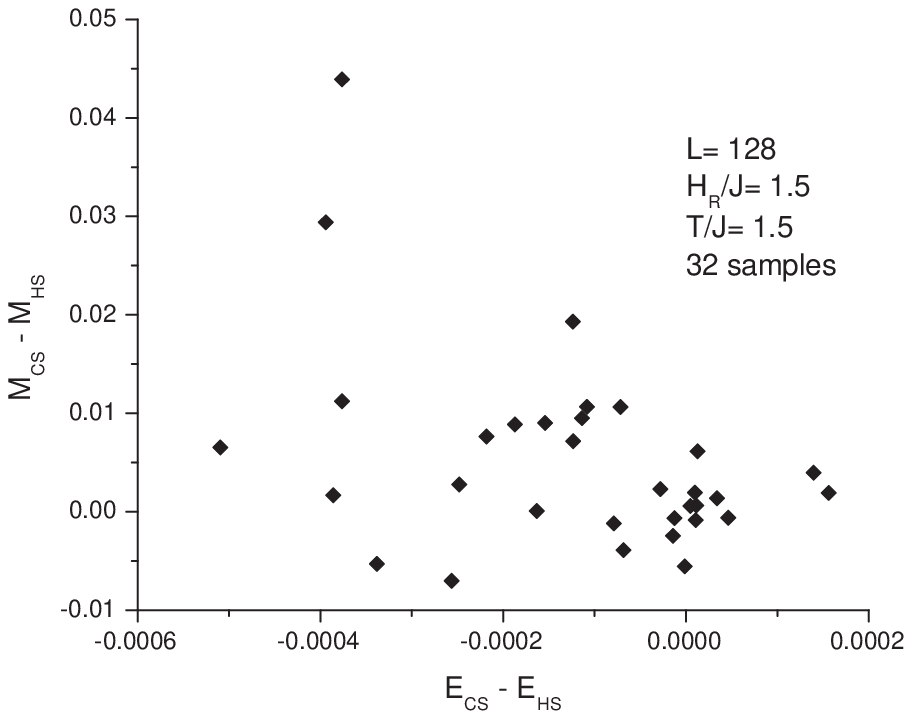}
\caption{\label{Fig.5} Jump in the magnetization vs. jump
in the energy for $128 \times 128 \times 128$ lattices with $h_r$ = 1.5
at $T$ = 1.5.  States with hot start and ordered cold start initial
conditions are compared for each sample.}
\end{figure}

\begin{figure}
\includegraphics[width=3.4in]{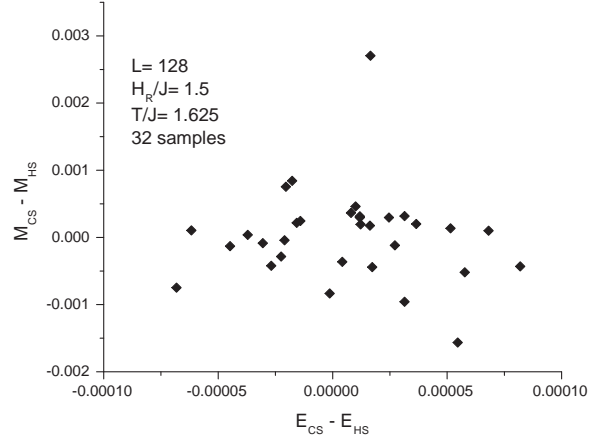}
\caption{\label{Fig.6} Jump in the magnetization vs. jump
in the energy for $128 \times 128 \times 128$ lattices with $h_r$ = 1.5
at $T$ = 1.625.  States with hot start and ordered cold start initial
conditions are compared for each sample.}
\end{figure}

We now return to the discussion of the $h_r = 1.5$ case.  The data for the
slowly warmed states and the slowly cooled states at the same value of $T$
are indistinguishable for all non-zero values of $|k|$.  However, this is
not necessarily true at $\vec{\bf k} = 0$ for a finite sample, as was
discussed in detail for the $h_r = 1.875$ case in ref.\cite{Fis19}. It is
not necessarily true that for a particular sample the spin state is very
similar for the warmed state and the cooled state.  What actually happens
for individual samples is that, in most cases the spin state of the slowly
warmed state with an ordered initial condition is significantly different
at $T = 1.421875$ from the slowly cooled state.  However, at $T = 1.625$
the slowly warmed state is, in most cases, essentially indistinguishable
from the slowly cooled state.  We illustrate this for $T = 1.421875$ in
Fig.~4, for $T = 1.5$ in Fig.~5 and for $T = 1.625$ in Fig.~6, which plot
the differences in the magnetization and the energy for individual samples.
Note that in most, but not all samples, at $T = 1.421875$ the warmed state
has a lower energy and a higher magnetization than the cooled state.  At
$T = 1.625$ the differences are much smaller, and they no longer have much
systematic dependence on the initial conditions.

In Table I we display data for the average magnetization per spin,
$|M(L)|/L^3$, the longitudinal magnetic susceptibility per spin,
$\chi_{||}/L^3$, and the specific heat at zero average field, $c_{H = 0}$.
It was found for $h_r = 1.875$ that $|M|$ appears to have a subextensive
divergence at $T_c$,\cite{Fis19} and it is expected that this will also
be true at $h_r = 1.5$.  However, $\lambda$ is somewhat longer
at $h_r = 1.5$.  Thus, in order to check how $|M(L)|$ scales with $L$ at
$h_r = 1.5$, we would need data for larger lattices, which is not practical
using the computers currently available.

\begin{quote}
\begin{flushleft}
Table I: Thermodynamic data for hot start and cold start initial
conditions at $h_r$ = 1.5, for various $T$. (hs) and (cs) mean
data obtained using hot start and cold start initial conditions,
respectively.  The one $\sigma$ statistical errors shown are due
to the sample-to-sample variations.
\begin{tabular}{|c|ccc|}
\hline
$T$&$|M|/L^3$&$\chi_{||}/L^3$&$c_{H = 0}$\\
\hline
1.421875{hs}&0.070$\pm$0.008&30.3$\pm$1.2&1.102$\pm$0.003\\
1.421875{cs}&0.102$\pm$0.008&28.8$\pm$1.1&1.097$\pm$0.003\\
1.5{hs}&0.051$\pm$0.005&33.9$\pm$1.3&1.204$\pm$0.004\\
1.5{cs}&0.056$\pm$0.005&33.6$\pm$1.4&1.120$\pm$0.004\\
1.625{hs}&0.028$\pm$0.003&25.2$\pm$0.5&1.326$\pm$0.004\\
1.75{hs}&0.0142$\pm$0.0012&12.96$\pm$0.36&1.302$\pm$0.004\\
1.875{hs}&0.0081$\pm$0.0006&6.71$\pm$0.22&1.147$\pm$0.004\\
\hline
\end{tabular}
\end{flushleft}
\end{quote}

There is a peak in $\chi_{||}/L^3$ centered close to $T = 1.5$, but it
appears to have a finite maximum, as was found for larger values of
$h_r$.\cite{Fis19}  As should be expected, the peak in $\chi_{||}/L^3$
increases in height as $h_r$ decreases.  Unless there is a phase transition
into a ferromagnetic phase, it is not expected that $\chi_{||}/L^3$ will
diverge to infinity for any $h_r \neq 0$.  There is a very broad peak in
$c_{H = 0}$ centered at about $T = 1.625$, which is not expected to be
associated with long-range correlations.  $T = 1.625$ is the temperature
where the thermal correlation length is equal to $\lambda$.
In the terminology of relaxor ferroelectrics, this is the Burns
temperature.\cite{BD83}

\section{Discussion}

The author thinks it is worth observing that the kind of jumps we are seeing in
the energy per spin and the magnetization per spin of finite samples would need to
disappear in the limit $T \to 0$. The  multicritical critical point hypothesis for
the behavior of random field models at $T$ = 0 says that $T$ should be an
irrelevant variable at that point.  However, the behavior we are seeing along the
phase transition line for $T > 0$ is not consistent with that hypothesis.  If RSB
creates a glassy phase\cite{BDD01} between the paramagnet and the ferromagnet when
$T > 0$, then this issue is resolved.  This is true for both the RFXYM and also the RFIM.

Finding that $S (\vec{\bf k})$ diverges at low temperatures in the RFXYM as
$|\vec{\bf k}| \to 0$ is not surprising.  This behavior follows from the results
of A. Aharony\cite{Aha78} for models which have a probability distribution
for the random fields which is not isotropic.  According to Aharony's
calculation, if this distribution is even slightly anisotropic, then we
should see a crossover to RFIM behavior at a sufficiently small value of $|k|$.
We know\cite{Imb84,BK87} that in $d = 3$ the RFIM is ferromagnetic at low
temperature if the random fields are not very strong.  The instability to even
a small anisotropy in the random field distribution should induce a diverging
response in $S (\vec{\bf k})$ as $|\vec{\bf k}| \to 0$ for the RFXYM in $d = 3$.
A similar effect in a related, but somewhat different, model was found by
Minchau and Pelcovits.\cite{MP85}.

More recently, models of quantum-mechanical spins in random fields have
been studied at $T = 0$.\cite{Car13,AN18}  These calculations find
logarithmic divergences of the structure factor as $|\vec{\bf k}| \to 0$
in these quantum versions of random field models.  It is not clear yet
that one should be able to map the classical RFXYM at finite temperature
onto a quantum model at $T$ = 0.  However, A. Aharony's argument about
the instability in the 3D RFXYM makes this connection plausible.

Note that it is only $S$ which diverges for the 3D RFXYM.  Unlike the
situation for the Kosterlitz-Thouless transition, we are not seeing any
divergence of $\chi$.  The difference in the behavior of $S$ and $\chi$
is due to the fact that the local magnetization, $\vec{\bf M_{i}}$, has
a non-zero average value even at high $T$ in a random field model.  What
is going on here is that the $Q_{ij}$ terms in Eqn.~4 are canceling
against the $\langle \cos ( \phi_{i} - \phi_{j} ) \rangle$ terms, and
giving a finite net result, even at $T_c$.  It is very unclear that the
behavior we are seeing can be attributed to topological defects.  However,
the range of uncertainty in $T_c$ is significant, and we cannot rule out that
the RSB phase transition occurs at the same temperature as the percolation
transition of the vortex lines on the dual lattice, as proposed by Gingras
and Huse.\cite{GH96}

Several years ago, calculations of Chudnovsky and coworkers\cite{GCP13,PGC14}
made much stronger claims.  These authors use a downhill relaxation algorithm
for the 3D RFXYM at $h_r = 1.5$.  The states found by their algorithm are
local energy minima of the Hamiltonian which have values of $|M|/L^3$ of
approximately 0.80.  We see no reason to believe that such a downhill
relaxation algorithm should be able to come anywhere close to finding the
true ground state of a sample for large $L$ at $h_r = 1.5$.  It is the current
author's opinion that in order for the results of such a downhill relaxation
algorithm to be convincing, they must be done using an $L$ which is a power
of 2.  In that case, $S (\vec{\bf k})$ could be calculated in the same way
it has been done here.  A properly relaxed state for a ferromagnetic state
of an $XY$ model must have a divergent peak of $S$ for $|k| \to 0$.

The results we are finding at $h_r = 1.5$ are qualitatively similar to the
results we found previously\cite{Fis19} at $h_r = 1.875$.  Chudnovsky
{\it et al.} say that they find no ground state magnetization near $h_r = 2.0$.
We consider an abrupt qualitative change in the ground state behavior
between $h_r = 1.5$ and $h_r = 2.0$, as claimed by Chudnovsky {\it et al.},
to be implausible for this model.  Since our Monte Carlo calculations are
limited to $L = 128$, we cannot obtain results in the regime where the thermal
correlation length is larger than $\lambda$ when $h_r \le 1.0$.  There has
been no attempt in this work to equilibrate samples at temperatures below
$T = 1.421875$.  Therefore, we have no data which directly address the question
of whether the RFXYM shows true ferromagnetism in $d = 3$.   We do not claim
that we know what happens for small values of $h_r$.

It appears to the author that what is going on in this model is a
broken ergodicity transition in the phase space, without any change
in the spatial symmetry.  In that sense, it is similar to a spin-glass
phase transition.  However, a random field model does not have the
two-fold Kramers degeneracy of a spin glass.  Therefore the broken
ergodicity occurs in the random field model in a purer form, without
the extra complication of the two-fold symmetry in the phase space.

The reader may be tempted to object that such a phase transition cannot
be described within the usual formalism of equilibrium statistical
mechanics, based on the canonical partition function
\begin{equation}
  Z ( T )~=~ {\bf Tr}_{\{ \phi_i \} } \exp ( - H / T ) \,  ,
\end{equation}
where $H$ is given in Eqn.~1.  We are thinking now about a particular
sample, so the $\theta_i$ variables are fixed.  For a classical system,
the standard formulas based on $Z$ do not have any dependence on
dynamics. That is the point.  The fact that our Monte Carlo calculation
sees that the hot start states and the cold start states we find for
$T \le 1.5$ are not the same means that these results cannot be
described by $Z ( T )$. Our calculation is not finding the partition
function.  When the dynamical relaxation time is infinite over a range
of $T$, $Z ( T )$ will not give us the behavior seen in a laboratory
experiment.  Of course, strictly speaking, the relaxation time is not
actually infinite in a finite sample.  However, real experiments are
done on finite samples, in finite times.

The idea of the broken ergodicity transition is exactly that we need to
include dynamics in order to understand what is going on.  It is true that
if we ran the Monte Carlo calculation for any finite lattice a very long
time, the results would, in principle, eventually converge to the $Z ( T )$
for that particular finite lattice.  However, there is an order of limits
issue.  A broken ergodicity transition, like all thermodynamic phase
transitions, only exists in the limit of an infinite system.  To get
correct results in the thermodynamic limit, we need to take the limit
$L \to \infty$ in an appropriate way.  We should not take the limit of
infinite time while holding $L$ fixed.  The results which come from a
Monte Carlo calculation may be thought of as telling us that the RSB in
the RFXYM is happening in three space dimensions and one time dimension
at some $T_c > 0$, if $h_r$ is not too large.  This is completely
independent of whether or not there might be a ferromagnetic transition at
some lower temperature.  A helpful review of Monte Carlo calculations,
which discusses critical slowing down of the dynamical behavior at a phase
transition, has been given by Sokal.\cite{Sok92}  One could say that, for
the RFXYM problem, dynamical slowing down is not a bug, it is a feature.

Hui and Berker\cite{HB89} argued that the vanishing of the latent heat
implied that a critical fixed point should exist.  This author does not
see, however, why such a fixed point, with its associated divergent
correlation length, should generally exist in a model which has no
translation symmetry, except in those cases where the randomness is an
irrelevant operator.\cite{Har74}  It is certainly true that there are
some cases where such fixed points have been found using $\epsilon$-expansion
calculations.  Subextensive singularities\cite{Fis19} in the specific heat and the
magnetization are completely consistent with the Aizenman-Wehr Theorem.\cite{AW89,AW90}

\section{Summary}

In this work we have performed Monte Carlo studies of the 3D RFXYM on $L = 128$
simple cubic lattices, with a random field strength of $h_r = 1.5$.  We compared
the properties of slowly cooled states and slowly heated states at $T = 1.421875$,
$T = 1.5$ and $T = 1.625$. The temperature at which there appears to be a phase
transition described by a divergence in the structure factor at $S (|\vec{\bf k}|
 = 0 )$ is probably between $T = 1.5$ and $T = 1.421875$.  The behavior is
qualitatively the same as what was found earlier\cite{Fis19} for somewhat larger
values of $h_r$.  We have also computed values of the magnetic susceptibility and
the specific heat.  The data are consistent with the idea that in $d = 3$ the
RFXYM has a phase transition into a phase described by broken ergodicity, as long
as the strength of $h_r$ is not too large.  We do not believe that there is a
ferromagnetic phase at any value of $T$ for $h_r = 1.5$.  These results appear to
be related to RSB\cite{BDD01}, and to recent work on quantum disorder.\cite{AN18}

\begin{acknowledgments}
The author thanks N. Sourlas for a helpful conversation about the recent work
on the random field Ising model, and Ofer Aharony for a discussion of his recent
results on quantum disordered models. This work used the Extreme Science and
Engineering Discovery Environment (XSEDE) Bridges Regular Memory at the Pittsburgh
Supercomputer Center through allocations DMR170067 and DMR180003.  The author
thanks the staff of the PSC for their help.

\end{acknowledgments}



\end{document}